\documentclass[12pt,letterpaper]{article}
\usepackage[utf8]{inputenc}

\usepackage{authblk} 
\usepackage{orcidlink}
\usepackage{graphicx,array}
\usepackage{url}
\usepackage{color}
\usepackage{latexsym}
\usepackage{amsthm}
\usepackage{amsmath}
\usepackage{amssymb}
\usepackage{amsfonts}
\usepackage[numbers,sort&compress]{natbib}
\usepackage{bm}
\usepackage{bbm}
\usepackage{slashed}
\usepackage{mathrsfs}
\usepackage{enumerate}
\usepackage{tikz}
\usepackage{siunitx}
\ExplSyntaxOn
\msg_redirect_name:nnn { siunitx } { physics-pkg } { none }
\ExplSyntaxOff
\usepackage{mdframed}
\usepackage{setspace}  
\usepackage{esvect}
\usepackage{physics}
\usepackage{enumitem}
\usepackage{booktabs}
\usepackage{tcolorbox}%

\numberwithin{equation}{section}

\usepackage{hyperref} 
\hypersetup{
    colorlinks=true,       
    linkcolor=red,          
    citecolor=blue,        
    filecolor=magenta,      
    urlcolor=blue           
}
\usepackage[all]{hypcap} 
\usepackage{multirow}
\usepackage{multicol}

\newcolumntype{M}[1]{>{\centering\arraybackslash}p{#1}}

\usepackage{natbib}
\setlist[description]{leftmargin=\parindent,labelindent=\parindent}

\usepackage[normalem]{ulem} 

\newcommand{\nc}{\newcommand}

\newcommand{\muB}{\mu_{\text{B}}}

\newcommand{\muR}{\mu_{\text{R}}}

\newcommand{\eg}{{\it e.g.}}
\newcommand{\ie}{{\it i.e.}}

\nc{\beq}{\begin{equation}}
\nc{\eeq}{\end{equation}}
\nc{\beqa}{\begin{eqnarray}}  
\nc{\eeqa}{\end{eqnarray}}  
\nc{\bit}{\begin{itemize}}  
\nc{\eit}{\end{itemize}}

\usepackage{floatrow}
\newfloatcommand{capbtabbox}{table}[][\FBwidth]

\usepackage{blindtext}

\allowdisplaybreaks

\title{ 
{\Large \bf QCD Vacuum Energy and Its Implication for\\ Quark Nugget Stability}
}

\author[a,b]{Yang Bai\orcidlink{0000-0002-2957-7319}}
\author[a]{Ting-Kuo Chen\orcidlink{0000-0002-5267-6729}}
\affil[a]{\small \it Department of Physics, University of Wisconsin-Madison, Madison, WI 53706, USA } 
\affil[b]{\small \it  HEP Division, Argonne National Laboratory, Argonne, IL 60439, USA} 

\date{}

\begin{document}

\maketitle

\setlength{\parskip}{0.2ex}

\begin{abstract}
In this work, we employ both theoretical and data-driven methods to derive the QCD vacuum energy, utilizing the GMOR relation, the low-energy theorem, and the equation of state from Lattice QCD. The QCD vacuum energy is determined to be between around $(163\,\mbox{MeV})^4$ and $(190\,\mbox{MeV})^4$. With the assumptions of complete deconfinement, vanishing gluon condensate, full chiral-symmetry restoration, and the validity of perturbative QCD at baryon chemical potentials of order of the proton mass, a very specific kind of quark nugget is found to be less stable than ordinary nuclei.
\end{abstract}

\thispagestyle{empty}  
\newpage    
\setcounter{page}{1}  

\begingroup
\hypersetup{linkcolor=black,linktocpage}
\tableofcontents
\endgroup

\newpage

\section{Introduction}

Quantum Chromodynamics (QCD) vacuum energy (or bag parameter) is one of the most fundamental quantities in QCD that dictates various phenomena of the physical universe. For example, it plays an important role in the study of (hybrid) neutron stars and quark stars whose existence and properties depend heavily on its value~\cite{Csaki:2018fls,Ventagli:2024cho}: As they consist of ordinary hadron matter and/or quark matter, the latter, if it actually exists given a dense enough stellar core, will exhibit a difference in its vacuum configuration from the former phase. Another example is the Cosmological Constant Problem: Given its dramatically lower current value of $(2.25\times 10^{-3}\,\mbox{eV})^4$~\cite{Planck:2018vyg} compared to the QCD scale ($\sim\mathcal{O}[(100~{\rm MeV})^4]$), it is crucial to more precisely determine the value of the vacuum energy associated with the QCD phase transition to explain this scale hierarchy. Finally, which is also our main example of applying the study on QCD vacuum energy, is its relation to the existence and stability of ``quark nuggets'', the solitonic ``bags'' of quark matter existing in the zero-pressure vacuum proposed by physicists such as Bodmer~\cite{Bodmer:1971we}, Lee and collaborators~\cite{Lee:1974kn,Lee:1974uu,Friedberg:1976eg,Friedberg:1977xf}, and Witten~\cite{Witten:1984rs}. To avoid confusion, we will use the term ``quark matter'' to refer to the collective quark degrees of freedom whether inside a stellar core or in a solitonic bag throughout the paper, while we reserve the term ``quark nugget'' for the solitonic bag of quark matter, although we will only study its bulk properties in the thermodynamic limit and neglect its surface and finite-size properties.

QCD vacuum energy is closely tied to the QCD trace anomaly that consists of two major parts: the gluon condensate and the quark bilinear condensates, the latter of which determine chiral symmetry. Because of its nonperturbative nature, determining the QCD vacuum energy is one of the most challenging parts of understanding QCD. To approach this issue, we analyze it from the following perspectives:
1. The Gell-Mann, Oakes, Renner (GMOR) relation~\cite{Gell-Mann:1968hlm}, which establishes a direct connection between the quark condensates and the pion mass and decay constant (also noticed in Ref.~\cite{Donoghue:2016tjk}).
2. The Novikov-Shifman-Vainshtein-Zakharov (NSVZ) low-energy theorem (LET) for QCD~\cite{Novikov:1981xi} and the measurement of QCD topological susceptibility $\chi_t$ through lattice QCD (LQCD)~\cite{Borsanyi:2016ksw}, which provides a robust lower bound on the gluonic contribution to the bag parameter.
3. The fitting to the isospin-dense LQCD (LQCD$_{\rm I}$)~\cite{Abbott:2024vhj} and finite-temperature LQCD (LQCD$_{\rm T}$) \cite{Borsanyi:2013bia,HotQCD:2014kol,Bazavov:2017dsy} equations of state based on the frameworks of pQCD and hot pQCD (HpQCD), respectively.
The former has already been conducted in Ref.~\cite{Bai:2024amm}, while the latter is presented in this study. Although the currently available LQCD calculations still cannot resolve the value of QCD vacuum energy due to data uncertainties, they can provide reliable upper bounds on it.
As we will show,  methods 1 and 2 can be directly used to scrutinize the stability of quark matter under certain assumptions which we will delineate later.

Stabilized by the balance between quark matter pressure and QCD vacuum pressure, quark nuggets might in fact be the global ground state of baryon matter in QCD instead of ordinary hadron matter. The study of the possible existence of quark nuggets is not only of theoretical importance to the exploration of QCD, but can also potentially lead to various phenomenological consequences. For instance, they could extend the periodic table by introducing ``exotic nuclei'' with very large atomic and atomic mass numbers~\cite{Holdom:2017gdc,Bai:2024muo}. Furthermore, if they were produced in the early universe and remain stable at the present time, they could serve as a compelling dark matter candidate~\cite{Witten:1984rs}.
While, as we will discuss later, one could in principle consider some more complicated scenarios that involve confinement and partial condensate restorations, we will assume complete deconfinement, zero gluon condensate, and complete chiral-symmetry restoration with zero quark bilinear condensates inside the quark nuggets, the same as in Ref.~\cite{Farhi:1984qu}. We emphasize that our later statement regarding stability is based on this strong assumption. We make no claim about the stability of quark nuggets in phases with partial deconfinement or chiral symmetry restoration.

For a given balanced quark nugget system, one can infer the associated energy per baryon $\epsilon/n_{\rm B}$. If $\epsilon/n_{\rm B}\leq 930$~MeV, the averaged mass per nucleon of the iron element (the most stable nucleus), then the quark nugget is stable in the thermodynamic limit (it will become less stable once finite-size effects are included since they will increase the total energy through surface tension, etc.).
Although it is commonly agreed that the energy scale of stable quark nugget is not necessarily high enough for it to be considered a fully perturbative system, pQCD is still a useful tool to investigate its thermodynamic properties without relying on phenomenological models (see Ref.~\cite{Holdom:2017gdc} for a study based on the linear sigma model with quarks). In the literature to date, the pQCD calculation of quark matter equation of state at finite baryon chemical potential has been carried out up to $\mathcal{O}(g^4)$ in Ref.~\cite{Kurkela:2009gj} and more recently partially addressed at $\mathcal{O}(g^6)$ in Refs.~\cite{Gorda:2021znl,Gorda:2021kme,Gorda:2023mkk,Karkkainen:2025nkz} ($g$ is the QCD gauge coupling). In the following, we will discuss how one can make use of the methods outlined previously to study QCD vacuum energy to approach the other side of the story.

This paper is organized as follows. In Section~\ref{sec:QCDVE}, we review the QCD vacuum energy and derive theoretical and/or data-driven bounds on it using the methods mentioned previously: the GMOR relation, the NSVZ low-energy theorem, and the fitting to LQCD equations of state, with the explicit HpQCD formulas for the quark-gluon plasma pressure listed in Appendix~\ref{sec:HpQCD}. In Section~\ref{sec:QM}, we discuss the existence of both stable 2+1- and 2-flavor quark nuggets and find that, under the assumptions of complete deconfinement and zero quark and gluon condensates, they are both excluded using $\mathcal{O}(g^4)$ pQCD calculations, the validity of which is discussed in Appendix~\ref{sec:NLO}. Finally, we conclude our study in Section~\ref{sec:conclusions}.

\section{QCD vacuum energy}\label{sec:QCDVE}

In the physical QCD vacuum with zero chemical potential and temperature, 
the QCD vacuum energy is related to the trace anomaly of QCD which results from the breaking of the classical dilatation symmetry at the quantum level. This trace anomaly operator is given by
\beqa\label{eq:trace_anomaly-main}
&\Theta^\mu_{\;\mu} &\equiv \Big[\frac{\beta(g)}{2g}G_a^{\mu\nu}G^a_{\mu\nu}
+ \sum_{q}\,m_q\,\gamma_m(g)\,\overline{q}q \Big]
 + \Big[\sum_{q}\,m_q\,\overline{q}q \Big]~.
\eeqa
Here, $\Theta^\mu_{\;\mu}$ is the trace anomaly operator that depends on the renormalized fields and the QCD gauge coupling $g$~\cite{Collins:1976yq,Nielsen:1977sy}, $q=u, d, s$ summing all three light quarks, $\beta(g)$ is the beta-function of the gauge coupling, and $\gamma_m(g)$ is the quark mass anomalous dimension~\cite{Hatta:2018sqd}. Note that the quantities inside each bracket are altogether renormalization-scale invariant. 

Usually, one defines the absolute value of the QCD vacuum energy as $-\frac{1}{4}$ times the nonperturbative contribution to the vacuum expectation value (VEV) of the trace anomaly in the physical vacuum, or $-\frac{1}{4}\langle \Theta^\mu_{\;\mu}\rangle_{0}$. 
The difference of VEV's of trace anomaly in the quark-gluon plasma phase and the physical vacuum can be defined as the positive bag parameter
$B^{\rm con}$, which contains two renormalization-scale invariant parts,
\beqa\label{eq:BGBF-main}
B^{\rm con}_G &=& -\frac{1}{4} \Big\langle \frac{\beta(g)}{2g}G_a^{\mu\nu}G^a_{\mu\nu} + \sum_q\,m_q\,\gamma_m(g)\,\overline{q}q \Big \rangle_0 \,, \\
B^{\rm con}_F &=& -\frac{1}{4} \Big\langle \sum_q\,m_q\,\overline{q}q \Big\rangle_0 ~.
\eeqa
The usual $B^{\rm con}$ is quoted to be $T$- and $\mu$-independent, \ie, at either high temperature or high chemical potential, the corresponding QCD quark-gluon plasma phase is fully (or approximately fully) deconfined, with zero gluon condensate and the chiral symmetry fully restored (the potential color superconducting phase in dense matter will be discussed later).
In the following, we use three different methods to obtain either partial or complete information (lower or upper bounds) of $B^{\rm con}_G$ and/or $B^{\rm con}_F$: the GMOR relation, the NSVZ low-energy theorem, and the fitting to LQCD equations of state.

\subsection{GMOR relation}\label{subsec:GMOR}

We begin our discussion of the QCD vacuum energy using the Gell-Mann, Oakes, Renner (GMOR) relation~\cite{Gell-Mann:1968hlm}, which was mentioned in Ref.~\cite{Donoghue:2016tjk} and presented in Ref.~\cite{Bai:2024amm}. Here, we simply summarize the results. For the $u, d$ light quarks, one has 
\begin{equation}
    \frac{f_\pi^2m_\pi^2}{2} = -m_\ell\langle\overline{\ell}\ell\rangle ,~ \ell=u,d ~,
\end{equation}
where $f_\pi\approx 92$~MeV is the pion decay constant, $m_\pi\approx 135$~MeV is the pion mass, and $m_\ell=(m_u+m_d)/2$ is the averaged light quark mass. Using the $\overline{\rm MS}$ scheme at 2~GeV, one gets that $\langle\overline{\ell}\ell\rangle^{\overline{\rm MS}}(2~{\rm GeV})=-(272
~{\rm MeV})^3$ based on the $SU(2)$ chiral perturbation theory~\cite{Borsanyi:2012zv}, as well as $m_u=2.16$~MeV, $m_d=4.70$~MeV, and $m_s=93.5$~MeV~\cite{ParticleDataGroup:2024cfk}, and thus   
\begin{equation}
    B^{\rm con}_{F,2}=-\frac{1}{4}\sum_{q=u,d}\langle m_q\overline{q}q\rangle^{\overline{\rm MS}}(2~{\rm GeV}) 
 = (76.8
 ~{\rm MeV})^4 ~.
\end{equation}
Correspondingly, the $s$-quark condensate $\langle\overline{s}s\rangle$ is further given by lattice measurement~\cite{McNeile:2012xh} and sum rule average~\cite{Narison:2004} respectively by~\cite{Bai:2024amm}
\begin{equation}\label{eq:s-condensate}
    \frac{\langle\overline{s}s\rangle^{\overline{\rm MS}}(2~{\rm GeV})}{\langle\overline{\ell}\ell\rangle^{\overline{\rm MS}}(2~{\rm GeV})} = \begin{cases}
        1.08\pm0.16 &~[\text{Lattice}] \\[1ex]
        0.66\pm0.10 &~[\text{Sum Rule Average}]
    \end{cases} ~,
\end{equation}
which leads to
\begin{equation}\label{eq:low_3-flavor}
    B^{\rm con}_{F, 2+1}\equiv -\frac{1}{4}\sum_{q=u,d,s}\langle m_q\overline{q}q\rangle^{\overline{\rm MS}}(2~{\rm GeV}) 
    = \begin{cases}
        (153\substack{+6.0 \\ -6.0}~{\rm MeV})^4 &~[\text{Lattice}] \\[1ex]
        (136\pm5.0~{\rm MeV})^4 &~[\text{Sum Rule Average}]
    \end{cases} ~.
\end{equation}
For later convenience, we will label the above values as GMOR$_{\rm L}$ and GMOR$_{\rm S}$, respectively. Consequently, these give $B^{\rm con}_{F,2}=(76.8
 ~{\rm MeV})^4$ and $B^{\rm con}_{F,2+1}=(153
 ~{\rm MeV})^4$ [GMOR$_{\rm L}$] or $B^{\rm con}_{F,2+1}=(136
 ~{\rm MeV})^4$ [GMOR$_{\rm S}$].

\subsection{Low-energy theorem}\label{subsec:LET}

In this subsection, we combine the Novikov-Shifman-Vainshtein-Zakharov (NSVZ) low-energy theorem (LET) and the LQCD calculation of topological susceptibility to derive a lower bound on $B^{\rm con}_{G}$. The NSVZ LET can be derived using the anomalous Ward identity of the dilatation symmetry~\cite{Novikov:1981xi}, which states that
\beqa
\label{eq:LET}
 \lim_{q\to0}i \int d^4x e^{iqx} \langle \mbox{T} \mathcal{O}(x) \,\Theta^\mu_{\,\mu}(0) \rangle_0 = (-d_{\mathcal{O}}) \,\langle \mathcal{O}\rangle_0 ~.
\eeqa
Here, $\mbox{T}$ stands for time ordering and  $d_{\mathcal{O}}$ is the canonical mass dimension of the operator $\mathcal{O}$, with $d_{\mathcal{O}}=4$ for $\mathcal{O} = G^{\mu\nu}_a G^{a}_{\mu\nu}$ and $d_{\mathcal{O}}=3$ for $\mathcal{O} = \bar{q}q$. To the leading order in small quark mass  and using the one-loop beta function for the gauge coupling, one has 
\begin{equation}\label{eq:LET-leading}
    \lim_{q\to0}i\int d^4x\, e^{iqx}\left\langle \mbox{T} \mathcal{O}(x)\, \frac{\alpha_s}{8\pi} G^2(0) \right\rangle_0 = \frac{d_{\mathcal{O}}}{\beta_0}\langle\mathcal{O}\rangle_0 + O(m_q) ~,
\end{equation}
where we have suppressed the Lorentz indices of the field strength tensors and simply write $G_{\mu\nu}^aG^{\mu\nu}_a\equiv G^2$. Here, $\beta_0=(11N_c-2N_f)/3$ is the coefficient of the leading term in $\beta(\alpha_s)\equiv d\alpha_s/d\log{\mu} = - \beta_0\alpha_s^2/(2\pi) + O(\alpha_s^3)$ [note that $\beta(g)\equiv dg/d\log{\mu} = - \beta_0\,g^3/(16\pi^2) + O(g^5)$]. For our purpose of estimating the gluonic contribution to the QCD vacuum energy, we ignore the $\mathcal{O}(m_q)$ corrections and make a more conservative estimation.

Applying the LET to the $G^2$ operator with $d_{\mathcal{O}} = 4$, one has
\beqa
i\int d^4x\, \left\langle \mbox{T} \frac{\alpha_s}{8\pi} G^2(x)\, \frac{\alpha_s}{8\pi} G^2(0) \right\rangle_0 = \frac{4}{\beta_0} \left\langle\frac{\alpha_s}{8\pi} G^2 \right\rangle_0 ~,
\eeqa
where we have used the equal sign by ignoring the small quark mass corrections. In the following, we will show that the left-hand side of the above equation can be related to the topological susceptibility through a simple algebraic inequality. We first translate this equation from the Minkowski spacetime to the Euclidean spacetime (note that $\tau = i t$, $G^2=-G^2_{\rm E}$ and $G\widetilde{G}=i\,G_{\rm E}\widetilde{G}_{\rm E}$ with $\widetilde{G}\equiv\widetilde{G}^{\mu\nu}=\frac{1}{2}\epsilon^{\mu\nu\rho\sigma}G_{\rho\sigma}$, where the fields without subscripts are defined in the Minkowski spacetime). Using the inequality
\begin{equation}
    G^2_{\rm E} = \frac{1}{2}(G^2_{\rm E}+\widetilde{G}^2_{\rm E}) = \frac{1}{2}\left[(G_{\rm E}-\widetilde{G}_{\rm E})^2+2\,G_{\rm E}\widetilde{G}_{\rm E}\right] \geq G_{\rm E}\widetilde{G}_{\rm E} ~,
\end{equation}
one can show that
\beqa
\frac{4}{\beta_0} \left\langle\frac{\alpha_s}{8\pi} G^2 \right\rangle_0 &=& \int d^4x_{\rm E}\, \left\langle \mbox{T} \frac{\alpha_s}{8\pi} G^2_{\rm E}(x)\, \frac{\alpha_s}{8\pi} G^2_{\rm E}(0) \right\rangle_0  \nonumber \\
\label{eq:chi_t}
&\ge& \int d^4x_{\rm E}\, \left\langle \mbox{T} \frac{\alpha_s}{8\pi} G_{\rm E}\widetilde{G}_E(x)\, \frac{\alpha_s}{8\pi} G_{\rm E}\widetilde{G}_E(0) \right\rangle_0 \equiv \chi_t  \\
\label{eq:BG-chi_t-bound}
\Rightarrow B^{\rm con}_G & \ge & \frac{\beta_0^2}{16} \chi_t ~.
\eeqa
In the last step, we ignored the positive $\gamma_m$-term in $B^{\rm cons}_G$, thus making the above lower bound more conservative.\footnote{Keeping the leading term in $\mathcal{O}(m_q)$, the bound becomes $B_G - \frac{1}{4}\langle \gamma_m m_q \bar{q} q\rangle \ge \frac{\beta_0^2}{16}\chi_t - \frac{3}{16}\langle m_q \bar{q} q\rangle$, where $\gamma_m\equiv -d\ln m_R(\mu)/d\ln \mu= 2\alpha_s/\pi + \mathcal{O}(\alpha_s^2)$ with $m_{R}(\mu)$ as the renormalized quark mass~\cite{Chetyrkin:1997dh,Hatta:2018sqd}. Note that since $\langle m_q \bar{q} q\rangle < 0$ and that $\gamma_m < 3/4$ for a perturbative $\alpha_s$, the bound becomes more stringent after including the leading term in $\mathcal{O}(m_q)$.}

The topological susceptibility, $\chi_t$, is a measure of the fluctuation of the topological charge $Q(x) = \frac{g^2}{32\pi^2}G_{\rm E}\widetilde{G}_{\rm E}(x)$ throughout spacetime. In the Euclidean spacetime, one simply has $\chi_t = \langle Q^2 \rangle/V$ with $V$ as the spacetime 4-volume. Based on chiral perturbation theory, one can relate $\chi_t$ to the pion mass and decay constant as $\chi_t = \frac{1}{4}(1+O[m_\pi^2/(4\pi f_\pi)^2])m_\pi^2 f_\pi^2$~\cite{Shifman:1979if,Mao:2009sy,Aoki:2009mx,Guo:2015oxa}.
Refs.~\cite{Borsanyi:2016ksw,Aoki:2017imx} have used LQCD to perform precise calculations of this quantity and reported results that are consistent with the prediction of the chiral perturbation theory. Here, we take the zero-temperature value in Ref.~\cite{Borsanyi:2016ksw}, $\chi_t =(75.6\pm 2.0\,\mbox{MeV})^4$, after combining the statistical and systematic errors. 

Substituting the value of $\chi_t$ into the bound given in Eq.~\eqref{eq:BG-chi_t-bound} and taking $\beta_0=9$ with $N_c=3$ and $N_f = 3$, we arrive at a robust lower bound on $B_G^{\rm con}$,
\beqa\label{eq:LET:result}
B^{\rm con}_G \ge \frac{\beta_0^2}{16} \chi_t = (113.4\,\mbox{MeV})^4 ~.
\eeqa
For clarity, the choice of $N_f=3$ comes from the number of flavors involved in the renormalization and has nothing to do with the flavors of quark matter actually existing in the system.

Before concluding, we mention another form of the LET that is related to $m_q$~\cite{Novikov:1981xi},
\begin{equation}\label{eq:LET_mq}
    \frac{d}{dm_q}\left\langle\frac{\alpha_s}{\pi}G^2\right\rangle_0 = -\frac{24}{\beta_0}\langle\overline{q}q\rangle_0 ~,
\end{equation}
which describes how the gluon condensate evolves with the light quark mass in the physical vacuum. However, it is unclear how this LET can be used to study the quark nugget for two reasons: (1) the most important effect of this should come from the $s$-quark, whose mass, however, is doubtfully light enough to validate this formula, and (2) even equipped with this formula, we cannot know the actual value of $\langle G^2\rangle_0$ since Eq.~\eqref{eq:LET_mq} only provides the derivative information of $\langle G^2\rangle_0$ with respect to $m_q$. Nevertheless, this formula does give us some hint about the possible partial restoration of the gluon condensate in the 2-flavor quark nugget system, which we will revisit later.

\subsection{Equation of state}\label{subsec:EOS}

The third way we infer about the QCD vacuum energy is by analyzing the equation of state calculated through LQCD at either finite temperature (LQCD$_{\rm T}$) or finite isospin density (LQCD$_{\rm I}$), assuming the validity of perturbation theories including pQCD and HpQCD. In this subsection, we present the details of our LQCD$_{\rm T}$ fit at zero quark chemical potential based on HpQCD, while the details of our LQCD$_{\rm I}$ study at nearly zero temperature [$T\sim\mathcal{O}(10)$~MeV]~\cite{Abbott:2024vhj} can be found in Ref.~\cite{Bai:2024amm}. 

We first recap our previous analysis of the LQCD$_{\rm I}$ data based on
the pQCD framework~\cite{Bai:2024amm}. By comparing the combined predictions of pQCD at finite density~\cite{Kurkela:2009gj} and the color superconducting (CS) gap~\cite{Fujimoto:2023mvc,Fujimoto:2024pcd} to the LQCD$_{\rm I}$ data from Ref.~\cite{Abbott:2024vhj} (earlier results can be found in Ref.~\cite{Abbott:2023coj}), we obtained the 90\%~CL upper bound on the two-flavor QCD vacuum energy $B_2^{\rm con}\lesssim(160~{\rm MeV})^4$. As we will discuss more later, it is plausible that the $s$-quark condensate remains nonzero and intact in the isospin-dense system, and thus the real 2+1-flavor QCD vacuum energy $B_{2+1}^{\rm con}$ can in principle be greater than this value.

To model the thermodynamic properties of a zero-density hot QCD system, we refer to the framework of HpQCD, which is an effective field theory formulated by integrating out various hard and soft thermal modes after performing dimensional reduction~\cite{Appelquist:1981vg,Nadkarni:1982kb}. The corresponding grand potential has been studied in the massless quark limit at orders 
$\mathcal{O}(g^2)$~\cite{Shuryak:1977ut,Chin:1978gj}, 
$\mathcal{O}(g^3)$~\cite{Kapusta:1979fh},
$\mathcal{O}(g^4\ln(1/g))$~\cite{Toimela:1982hv},
$\mathcal{O}(g^4)$~\cite{Arnold:1994ps,Arnold:1994eb},
$\mathcal{O}(g^5)$~\cite{Zhai:1995ac}, and
$\mathcal{O}(g^6\ln(1/g))$~\cite{Kajantie:2002wa}, while the massive quark effects have been studied up to $\mathcal{O}(g^2)$~\cite{Laine:2006cp}. It is long known that nonperturbative effects kick in at $\mathcal{O}(g^6)$, which result from the infinite number of diagrams contributing to the grand potential through the gluon magnetic mass at the same order~\cite{Linde:1980ts,Gross:1980br}. Although some of these nonperturbative terms have been studied in the literature~\cite{Hietanen:2004ew,DiRenzo:2006nh}, including the most recent Ref.~\cite{Navarrete:2024ruu}, the full expression at $\mathcal{O}(g^6)$ remains unknown. To address this issue, we follow the common prescription of assigning a dimensionless parameter $\Delta$ at $\mathcal{O}(g^6)$ to be fitted from the lattice data, assuming that it can successfully describe the nonperturbative effects.\footnote{
The argument for the validity of this prescription can be found in Ref.~\cite{Kajantie:2002wa}, which relies on the $T$- and $\overline{\mu}$-dependence of the unknown $\mathcal{O}(g^6)$ terms.
} 
As we will show later, this prescription can describe the lattice data very efficiently, though one cannot know at this point the correctness of the parameter choice. 

The quantity that we choose to fit is the normalized trace anomaly,
\begin{equation}
    \frac{\Theta^\mu_{~\mu}}{T^4} = T\frac{\partial}{\partial T}\left(\frac{p_{\rm QCD}}{T^4}\right) = T\frac{\partial}{\partial T}\left(\frac{p_{\rm QCD,m}}{T^4}\right) + \frac{4B_{2+1}^{\rm con}}{T^4} ~,
\end{equation}
where the HpQCD formulas for the total QCD pressure $p_{\rm QCD}$ and the quark-gluon plasma pressure $p_{\rm QCD,m}$ are summarized in Appendix~\ref{sec:HpQCD}. 
Note that here $\Theta^\mu_{~\mu}$ stands for the VEV of the trace anomaly operator in the finite-temperature vacuum in accordance with the LQCD notation and should not be confused with the operator defined earlier.
Here, we have chosen to model the QCD vacuum energy using the constant value $B_{2+1}^{\rm con}$, which is only asymptotically valid at high enough $T$ given the crossover nature of the QCD phase transition in the normal QCD vacuum~\cite{Aoki:2006we,Bhattacharya:2014ara}. Otherwise, one might need to consider additional temperature-dependent bag parameters, which do not directly contribute to the zero-temperature quark nugget properties but may affect the fit to the LQCD data.
Note that after fixing the number of flavors $N_f=2+1$, one has $p_{\rm QCD,m}=p_{\rm QCD,m}(X_T,T)$, where $X_T\equiv\overline{\mu}/(2\pi T)$ is the renormalization scale parameter with $\overline{\mu}$ being the renormalization scale of the system. 
The reason for this parametrization, as we have argued in Ref.~\cite{Bai:2024amm}, is because of the uncertainty in the choice of $\overline{\mu}$ with respect to the actual physical scale of a given system (see Refs.~\cite{Schneider:2003uz,Kurkela:2009gj} for example).
As a result, there are three parameters to fit from the LQCD$_{\rm T}$ data: $X_T$, $\Delta$, and $B^{\rm con}_{2+1}$. 
Since one should expect the confined plus chiral-symmetry-broken phase to have a lower vacuum energy than the deconfined and chiral-symmetry-restored phase, we require a priori that $B_{2+1}^{\rm con}\geq 0$.

\begin{figure}[th!]
    \centering
    \includegraphics[width=0.7\linewidth]{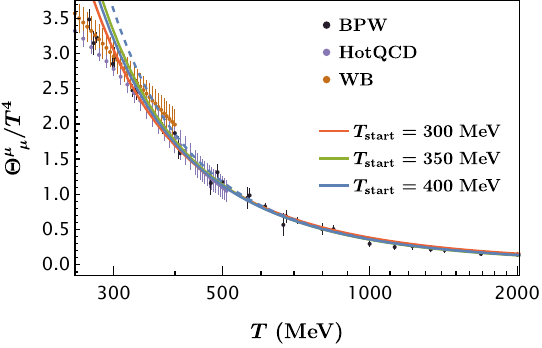}
    \caption{The lattice trace anomaly data of the zero-density 2+1-flavor hot QCD (LQCD$_{\rm T}$) system from Refs.~\cite{Borsanyi:2013bia,HotQCD:2014kol,Bazavov:2017dsy}, as well as the best-fit curves (solid) with $T_{\rm start}=300,350,400$~MeV. For illustration purposes, we also show the dashed line which is for the $T_{\rm start}=400$~MeV best-fit point but with a nonzero $B_{2+1}^{\rm con}=(170~{\rm MeV})^4$. Note that the errorbars of the LQCD$_{\rm T}$ data contain both systematic and statistical uncertainties.
    }
    \label{fig:LQCD_trace_anomaly-main}
\end{figure}

On the other hand, the LQCD calculations of the zero-density 2+1-flavor hot QCD system had been carried out up to $T\sim400$~MeV by the Wuppertal-Budapest (WB) collaboration~\cite{Borsanyi:2013bia} and $\sim500$~MeV by the HotQCD collaboration~\cite{HotQCD:2014kol}. Later on, Ref.~\cite{Bazavov:2017dsy} (BPW) pushed the calculation further up to $T\sim2$~GeV. More recently,  Ref.~\cite{Bresciani:2025vxw} has performed the simulations for $N_f=3$ massless quarks and $T\sim3$--$165$~GeV. Since the bag parameter is only relevant in the low-$T$ regime, we do not include this latest dataset in our current study. In principle, one should expect the HpQCD and LQCD$_{\rm T}$ calculations to match only when $T$ is considerably high where one can trust the perturbative calculations. Therefore, we truncate the data by introducing a starting temperature $T_{\rm start}$ and only fit the data with $T\geq T_{\rm start}$. Nevertheless, as we will demonstrate below, the $\Delta$ prescription renders a pretty good fit even down to $T_{\rm start}=300$~MeV. We collect the trace anomaly data measured in these works in Figure~\ref{fig:LQCD_trace_anomaly-main}, where the errorbars contain both systematic and statistical uncertainties.

We choose $T_{\rm start}=300,350,400$~MeV, and the best-fit points as well as the corresponding degrees of freedom $k$ and $\chi^2$ values are given by 
\begin{equation}
\begin{aligned}
    T_{\rm start} = 300~{\rm MeV}:&~\quad (k,\chi^2,X_T,\Delta,B_{2+1}^{\rm con}) = (71,55.8,1.65,-3338,0) ~, \\
    T_{\rm start} = 350~{\rm MeV}:&~\quad (k,\chi^2,X_T,\Delta,B_{2+1}^{\rm con}) = (52,23.6,1.52,-3087,0) ~, \\
    T_{\rm start} = 400~{\rm MeV}:&~\quad (k,\chi^2,X_T,\Delta,B_{2+1}^{\rm con}) = (35,13.2,1.56,-3155,0) ~,
\end{aligned}
\end{equation}
with the corresponding best-fit curves shown in Figure~\ref{fig:LQCD_trace_anomaly-main}. 
We note that the data could prefer $B_{2+1}^{\rm con}<0$, which we do not take into account since it is unphysical as stated before. One can see that all the curves describe the LQCD$_{\rm T}$ data very well, even in the high-$T$ regime where the uncertainties are relatively small. On the other hand, because of the large uncertainties in the low-$T$ regime where $B_{2+1}^{\rm con}$ plays a more significant role, the value of $B_{2+1}^{\rm con}$ remains unresolvable with the current LQCD$_{\rm T}$ measurements. For comparison, we also show the curve predicted by the $T_{\rm start}=400$~MeV best-fit point but with $B_{2+1}^{\rm con}=(170~{\rm MeV})^4$, which still agrees well with the high-$T$ data but clearly deviates from the low-$T$ data more compared to the original best-fit curve.

\begin{figure}[th!]
    \centering
    \includegraphics[width=0.48\linewidth]{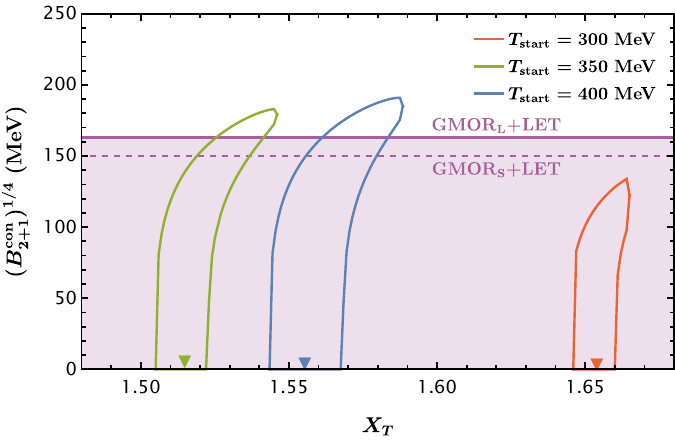} \hspace{3mm}
    \includegraphics[width=0.48\linewidth]{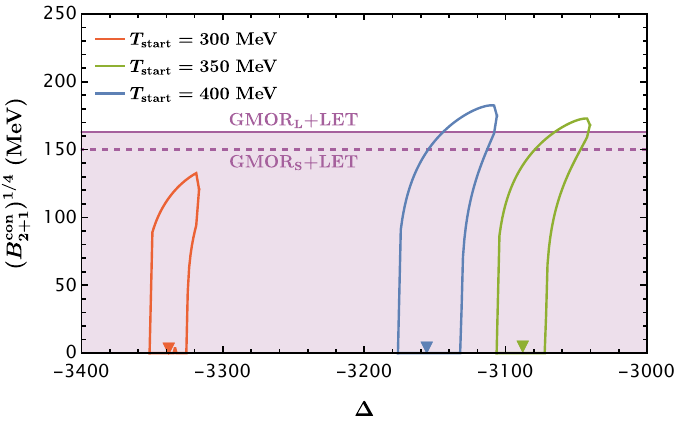}
    \caption{{\it Left panel:} The two-parameter 90\%~CL contours on the $[X_T,(B_{2+1}^{\rm con})^{1/4}]$ plane for $T_{\rm start}=300,350,400$~MeV. The horizontal lines are the lower bounds on $B_{2+1}^{\rm con}$ derived from the GMOR relation using either the lattice (GMOR$_{\rm L}$) or the sum rule average (GMOR$_{\rm S}$) [see Eq.~\eqref{eq:low_3-flavor}] plus the low-energy theorem (LET) [see Eq.~\eqref{eq:LET:result}]. {\it Right panel:} Same as the left panel but on the $[\Delta,(B_{2+1}^{\rm con})^{1/4}]$ plane.
    }
    \label{fig:contour}
\end{figure}

To obtain the constraints on the model parameters, we present the two-parameter 90\%~confidence level (CL) contours on the $[X_T,(B_{2+1}^{\rm con})^{1/4}]$ and $[\Delta,(B_{2+1}^{\rm con})^{1/4}]$ planes in Figure~\ref{fig:contour}. We also show the lower bounds on $B_{2+1}^{\rm con}$ based on the values of the quark condensates derived from the GMOR relation (see Section~\ref{subsec:GMOR}) and the low-energy theorem (LET) (see Section~\ref{subsec:LET})
in Figure~\ref{fig:contour}. In both parameter spaces, the upper bounds inferred from the $T_{\rm start}=300$~MeV case are in tension with the lower bounds derived before, while for the $T_{\rm start}=350$ and $400$~MeV cases, the loosest bound $B_{2+1}^{\rm con}\lesssim(190~{\rm MeV})^4$ shows up when we fix the best-fit $\Delta$ for $T_{\rm start}=400$~MeV. Consequently, we arrive at the following upper bound on the QCD vacuum energy based on the 2+1-flavor hot QCD system
\begin{equation}
   B^{\rm con}_{2+1} \lesssim (190~{\rm MeV})^4 ~,\qquad \mbox{(from LQCD}_{\rm T}\mbox{)}~,
\end{equation}
which is consistent with the bounds obtained from the pQCD+LQCD$_{\rm I}$+CS analysis if we assume that the $s$-quark condensate remains nonzero and intact in the isospin-dense system. We also note that although setting $T_{\rm start}$ to an even higher temperature than $400$~MeV will lead to a much looser upper bound on $B_{2+1}^{\rm con}$, we have shown that the consistency between the HpQCD predictions and lattice data for $T\gtrsim450$~MeV is pretty much unaffected by the different choices of $B_{2+1}^{\rm con}$. While this does not affect our subsequent analysis on quark nugget properties, which cares about the lower bound on the vacuum energy, we decide to quote the results obtained with $T_{\rm start}=400$~MeV to produce a meaningfully lenient bound that is consistent with the other theoretical bounds.

Subtracting the $s$-quark condensate contribution from the total vacuum energy, one obtains the upper bound on $B_2^{\rm con}$ for the two light flavor system as
\begin{equation}
    B_2^{\rm con} \lesssim \begin{cases}
        (168~{\rm MeV})^4 &~[\text{Lattice}] \\[1ex]
        (178~{\rm MeV})^4 &~[\text{Sum Rule Average}]
    \end{cases} ~,\qquad \mbox{(from LQCD}_{\rm T}\mbox{)} ~.
\end{equation}
Further subtracting the light-quark condensate contributions, one then has the upper bound on $B_{G}^{\rm con}$ as
\begin{equation}
    B_{G}^{\rm con} \lesssim \begin{cases}
        (166~{\rm MeV})^4 &~[\text{Lattice}] \\[1ex]
        (176~{\rm MeV})^4 &~[\text{Sum Rule Average}]
    \end{cases} ~,\qquad \mbox{(from LQCD}_{\rm T}\mbox{)} ~.
\end{equation}
For the following study on quark nugget stability, we will only use the upper bound derived from LQCD$_{\rm I}$ for $B_2^{\rm con}$ and that from LQCD$_{\rm T}$ for $B_{2+1}^{\rm con}$ so as to match the LQCD simulation setups with the corresponding directly inferred upper bounds:
\begin{equation}
    B_{2}^{\rm con} \lesssim (160~{\rm MeV})^4 ,~\qquad\qquad   B_{2+1}^{\rm con} \lesssim (190~{\rm MeV})^4 ~.
\end{equation}

\subsection{Summary: bounds on the QCD vacuum energy}

We briefly summarize the previous results.
For the QCD vacuum energy (VE) defined as $B_{2+1}^{\rm con}$, we have the lower bound from the GMOR relation plus the lattice (GMOR$_{\rm L}$)/sum rule average  (GMOR$_{\rm S}$) derivation for the strange quark condensate (Section~\ref{subsec:GMOR}) plus the low-energy theorem (LET) (Section~\ref{subsec:LET}) and the upper bound from the equation of state (Section~\ref{subsec:EOS}): 
\beqa\label{eq:bounds-main}
    \begin{cases}
        (163~{\rm MeV})^4~[\text{GMOR}_\text{L}\text{+LET}] \\[1ex]
        (150~{\rm MeV})^4~[\text{GMOR}_\text{S}\text{+LET}]
    \end{cases} \hspace{-0.3cm} < {\rm QCD~VE}\equiv B_{2+1}^{\rm con} < (190~{\rm MeV})^4~[\text{LQCD}_\text{T}\text{+HpQCD}] ~.
\eeqa
For the 2-flavor quark nugget property, we have also the following partial QCD vacuum energy when the strange quark condensate is nonzero and intact with
\beqa\label{eq:bounds-2-flavor-main}
    (119~{\rm MeV})^4~[\text{GMOR+LET}]~ < B_2^{\rm con} < (160~{\rm MeV})^4~[\text{LQCD}_\text{I}\text{+pQCD+CS}] ~,
\eeqa
where CS stands for color superconductivity.

\section{Quark nugget stability}\label{sec:QM}

As mentioned earlier, we assume complete deconfinement and zero gluon and chiral condensates within the quark nugget systems in this study, which is the only scenario that can be described by pQCD. Therefore, our inference about quark-matter stability applies only to this specific scenario and does not extend to quark nuggets with partially restored confinement and chiral symmetry.

\begin{figure}[t!]
    \centering
    \includegraphics[width=0.7\linewidth]{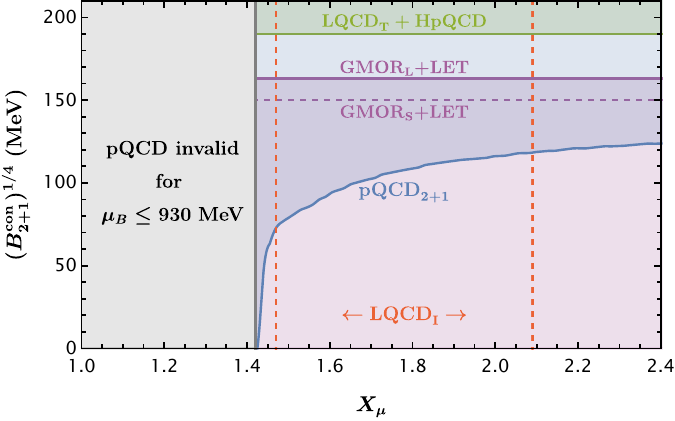}
    \caption{The $\epsilon/n_{\rm B}=930$~MeV contour (blue) predicted by the 2+1-flavor pQCD (pQCD$_{2+1}$) on the $[X_\mu,(B_{2+1}^{\rm con})^{1/4}]$ plane~\cite{Bai:2024amm} with $X_\mu$ defined in Eq.~\eqref{eq:X-param:main}, and stable 2+1-flavor quark nugget is excluded in the blue shaded region. The horizontal green and purple lines show the bounds on $B^{\rm con}_{2+1}$ given in Eq.~\eqref{eq:bounds-main}, the shaded regions above and below which denote the excluded value ranges of $B^{\rm con}_{2+1}$, respectively. pQCD is invalid for $\mu_{\rm B}<930$~MeV in the gray shaded region. The region between the two vertical red dashed lines marks the favored $X_\mu$ range by the LQCD$_{\rm I}$ data analyzed in Ref.~\cite{Bai:2024amm}.
    }
    \label{fig:X_bag_3-main}
\end{figure}

\begin{figure}[t!]
    \centering
    \includegraphics[width=0.7\linewidth]{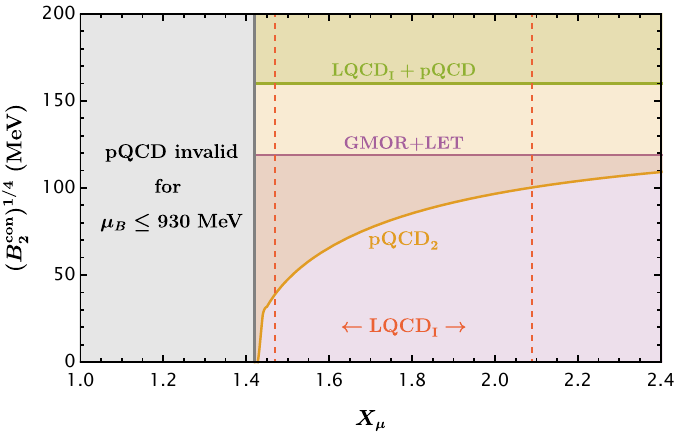}
    \caption{Similar to Figure~\ref{fig:X_bag_3-main} but for the $\epsilon/n_{\rm B}=930$~MeV contour (orange) predicted by the 2-flavor pQCD (pQCD$_{2}$) on the $[X_\mu,(B_2^{\rm con})^{1/4}]$ plane~\cite{Bai:2024amm}. The shaded regions are excluded by the mentioned respective constraints in a similar fashion to those in Figure~\ref{fig:X_bag_3-main}. 
    }
    \label{fig:X_bag_2-main}
\end{figure}

We follow the $\mathcal{O}(g^4)$ pQCD formalism~\cite{Kurkela:2009gj} to derive the quark matter equation of state as in Ref.~\cite{Bai:2024amm}
and leave the detailed discussion of its validity to Appendix~\ref{sec:NLO}. Here we simply note that the general trend of the $\mathcal{O}(g^6)$ correction seems to support our conclusions~\cite{Gorda:2021znl,Gorda:2021kme,Gorda:2023mkk} up to a still unknown constant to be determined in the future~\cite{Karkkainen:2025nkz}. 
We have also shown in Ref.~\cite{Bai:2024amm} that the CS gap contribution to the quark matter pressure is negligible using the next-to-leading-order formulas given in Refs.~\cite{Fujimoto:2023mvc,Fujimoto:2024pcd}, which give the color-flavor-locked (CFL) phase gap $\Delta_{\rm CFL}\approx4.5$~MeV for $\mu_{\rm B}=1$~GeV, leading only to an $\mathcal{O}(10^{-4})$ correction to the quark matter pressure, although the observational constraint on its actual value is still fairly loose (see for example Ref.~\cite{Geissel:2025vnp}, which suggests $\Delta_{\rm CFL}\lesssim\mathcal{O}(200)$~MeV at $\mu_{\rm B}=1$~GeV using neutron star conditions).
When the quark matter pressure and the vacuum pressure are balanced, \ie, $p_{\rm QCD}(\mu_{\rm B})-B^{\rm con}=0$, one can derive the corresponding energy per baryon $\epsilon/n_{\rm B}=\mu_{\rm B}$ of the (electrically neutral) quark nugget in the thermodynamic limit, which is in general a function of $B^{\rm con}$, $N_f$, and the renormalization scale parameter $X_\mu$ defined with respect to the quark chemical potentials as
\begin{equation}\label{eq:X-param:main}
    X_\mu = \begin{cases}
        \dfrac{\muR}{\frac{2}{N_c}\left(\mu_u+\mu_d+\mu_s\right)} &~ \text{[2+1-flavor]} \\[2ex]
        \dfrac{\muR}{\frac{2}{N_c}\left(\mu_u+2\mu_d\right)} &~ \text{[2-flavor]} \\[2ex]
        \dfrac{\muR}{\mu_u-\mu_d} &~ \text{[isospin-dense]}
    \end{cases} ~,
\end{equation}
$\muR$ being the renormalization scale. 
As we have studied in Ref.~\cite{Bai:2024amm}, the stable 2+1-flavor quark nugget is already excluded by considering the lower bound imposed by the GMOR relation under the given assumptions. For demonstration purposes, we show the corresponding $\epsilon/n_{\rm B} = 930$~MeV contour in Figure~\ref{fig:X_bag_3-main} together with the bounds on $B^{\rm con}_{2+1}$ from Eq.~\eqref{eq:bounds-main}.
On the other hand, the fate of the stable 2-flavor quark nugget still relies on the value of $B_{G}^{\rm con}$ if we assume that the $s$-quark condensate remains intact. We show in Figure~\ref{fig:X_bag_2-main} the contour of $\epsilon/n_{\rm B}=930$~MeV predicted by the 2-flavor pQCD (pQCD$_{2}$) in Ref.~\cite{Bai:2024amm} and the bounds given in Eq.~\eqref{eq:bounds-2-flavor-main}.

In both plots, the gray region with $X_\mu \lesssim 1.4$ denotes where pQCD is unreliable since $n_{\rm B}(\muB)<0$ when $\muB<930$~MeV~\cite{Kurkela:2009gj,Bai:2024amm} and thus one may not be able to draw a robust conclusion. However, given the monotonically increasing behavior of pQCD$_{2,2+1}$ as a function of $X_\mu$ and the large allowed value of $B^{\rm con}_{2,2+1}$, the balanced quark nugget system should prefer $X_\mu>1.4$.  On the other hand, though we do not show on the plot, the pQCD$_{2}$ curve actually intersects with the GMOR+LET curve at $X_\mu=3.10$, and thus the 2-flavor quark nugget can still be stable for $X_\mu\geq 3.10$. However, the existence of a system with such a large renormalization scale compared to the physical scale is questionable, as supported by its inconsistency with the preferred range of $1.47 < X_\mu < 2.09$ (vertical dashed red lines) by the LQCD$_{\rm I}$ data analyzed in Ref.~\cite{Bai:2024amm}. If we approximate $X_\mu\approx\mu_{\rm R}/[(2/3)\mu_{\rm B}]$ (which is exact in the massless limit), this range corresponds to $\mu_{\rm R}\in[911.4,1296]$~MeV for $\mu_{\rm B}=930$~MeV. For $X_\mu\leq1.5$ (and thus $\mu_{\rm R}\leq930$~MeV), the exclusion contours shown in Figures~\ref{fig:X_bag_3-main} and \ref{fig:X_bag_2-main} lie much below the lower bounds on QCD vacuum energy, and thus the conclusion is fairly certain, while for $X_\mu>1.5$ (and thus $\mu_{\rm R}>930$~MeV), the calculations should be trustworthy.

\begin{figure}[th!]
    \centering
    \includegraphics[width=0.48\linewidth]{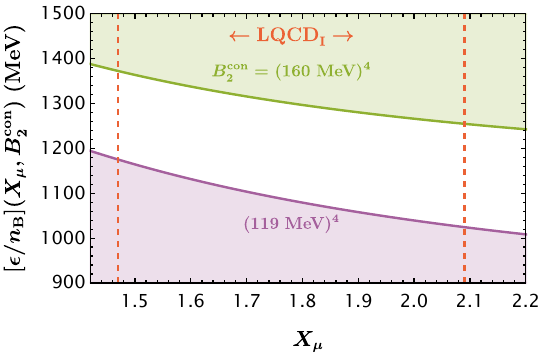}
    \includegraphics[width=0.48\linewidth]{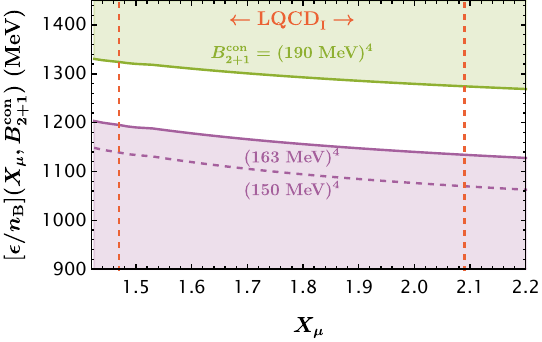}
    \caption{{\it Left panel:} The 2-flavor quark nugget mass per baryon or $[\epsilon/n_{\rm B}](X,B^{\rm con}_2)$ as a function of $X_\mu$ for $B_2^{\rm con}=(119~{\rm MeV})^4$ and $B_2^{\rm con}=(160~{\rm MeV})^4$, corresponding to the bounds given in Eq.~\eqref{eq:bounds-2-flavor-main}. 
    The region between the two vertical red dashed lines denotes the combined favored $X_\mu$ range by the LQCD$_{\rm I}$ data~\cite{Bai:2024amm}. 
    {\it Right panel: } Same as the left panel but for the 2+1-flavor quark nugget mass with $B_{2+1}^{\rm con}=(150~{\rm MeV})^4$, $(163~{\rm MeV})^4$, and $(190~{\rm MeV})^4$, corresponding to the bounds given in Eq.~\eqref{eq:bounds-main}.
    }
    \label{fig:EA_X_23}
\end{figure}

In the left panel of Figure~\ref{fig:EA_X_23}, we show the predicted quark nugget mass per baryon or  $\epsilon/n_{\rm B}$ curves for $B_2^{\rm con}=(119~{\rm MeV})^4$ and $B_2^{\rm con}=(160~{\rm MeV})^4$, corresponding to the lower and upper bounds given in Eq.~\eqref{eq:bounds-2-flavor-main}. Together with the favored range $1.47<X_\mu<2.09$ based on the LQCD$_{\rm I}$ data~\cite{Bai:2024amm}, we arrive at the following range for the 2-flavor quark nugget mass per baryon 
\beqa
\qquad 1020\,\mbox{MeV} < \frac{\epsilon}{n_{\rm B}} < 1370\,\mbox{MeV} ~, \qquad (\mbox{2-flavor quark nugget})~,
\eeqa
which clearly shows that the 2-flavor quark nugget is less stable than a nucleon. 

For the $N_f=2+1$-flavor quark nugget, we show the corresponding $\epsilon/n_{\rm B} = 930$~MeV contour in Figure~\ref{fig:X_bag_3-main} together with the bounds on $B^{\rm con}_{2+1}$ from Eq.~\eqref{eq:bounds-main}. Similar to the 2-flavor quark nugget case, the stable 2+1-flavor quark matter is also excluded. In the right panel of Figure~\ref{fig:EA_X_23}, we show the mass per baryon for the 2+1-flavor quark nugget as a function of $X_\mu$. Within the preferred range of $X_\mu$ from analyzing the LQCD$_{\rm I}$ data, the 2+1-flavor quark nugget has mass per baryon of
\beqa
\qquad\quad  1130\,\mbox{MeV} < \frac{\epsilon}{n_{\rm B}} < 1320\,\mbox{MeV} ~, \qquad (\mbox{2+1-flavor quark nugget})~,
\eeqa
which is also heavier than the nucleon mass. 

We emphasize again that the previous arguments were made based on the assumptions that the gluon and quark condensates are completely restored in the quark matter phase or that there is a first-order phase transition when $\mu_{\rm B}$ is above some critical value~\cite{Kurkela:2009gj,Rodrigues:2011zza,Fraga:2013qra,Xia:2017epi,Sedaghat:2024rjk,Fujimoto:2024pcd,Restrepo:2025qgp}. 
In reality, this might not necessarily be true in a dense environment, as suggested in Refs.~\cite{Novikov:1981xi,Shuryak:1980tp,Shuryak:1982hk}. With the partial restoration of the condensates, there is a chance that the resulted ``quark nugget'' (not the quark nugget historically defined with the complete restoration of the condensates in, \eg, Ref.~\cite{Farhi:1984qu}) might become stable under the reduced vacuum pressure.
Nevertheless, we are not able to rely on the existing pQCD to calculate the corresponding thermodynamic properties in this case and to obtain a conclusive result, since pQCD is only well established in the completely deconfined and zero-condensate scenario. For more complicated scenarios with confinement and partial condensate restoration, one has to go beyond the methods used in this study. 

\section{Discussion and conclusions}\label{sec:conclusions}

Apart from the potential alleviation coming from the nonperturbative contributions to the quark matter and QCD vacuum pressure, another possible mechanism to stabilize the quark nuggets is to introduce new physics that can either supply additional charges beyond the baryon number to account for stability or an additional order-parameter potential that changes the energy configuration of the quark nugget. Some examples are a scalar field that interacts with the topological sector of QCD and leads to the formation of domain walls (see for instance Ref.~\cite{Bai:2023cqj}) and the ``axion quark nuggets''~\cite{Zhitnitsky:2002qa}. These new physics are well motivated not only for their potential contribution to the stability of the quark nuggets, but also for the possible production mechanism of the quark nuggets in the early universe that they can provide, in contrast to the smooth QCD crossover in the Standard Model QCD~\cite{Aoki:2006we,Bhattacharya:2014ara}.

In addition to the stability of quark nugget, the QCD vacuum energy between $(163\,\mbox{MeV})^4$ and $(190\,\mbox{MeV})^4$ is also an important factor in the study of (hybrid) neutron stars, quark stars, and the Cosmological Constant (CC) Problem~\cite{Weinberg:1988cp,Polchinski:2006gy,Donoghue:2016tjk}. For a neutron star with a dense enough core, it is expected that the core matter will exist in the form of quarks. As the transition from ordinary hadron matter to quark matter in a cold baryon-dense environment is expected to be first-order, 
the 
associated vacuum energy will determine the properties of the hybrid stars (see for example Refs.~\cite{Csaki:2018fls,Ventagli:2024cho}). 
Similarly, the existence and properties of quark stars also highly rely on the value of the QCD vacuum energy.
On the other hand, the CC in the current universe is $(2.25\times 10^{-3}\,\mbox{eV})^4$~\cite{Planck:2018vyg}, which is $(2.0, 3.6)\times 10^{-44}$ of the QCD vacuum energy bounds derived in this study. Given that the QCD phase transition is the last known phase transition to modify the vacuum energy, any dynamical solution to the CC problem is likely to be concerned with QCD and possibly the precise value of the QCD vacuum energy. 

In conclusion, the nonperturbative QCD vacuum energy has been studied via the VEV of the QCD trace anomaly, with a focus on the completely deconfined, zero gluon-condensate, and chiral-symmetric phase that is parametrized by the $T$- and $\mu$-independent $B^{\rm con}$.
We have adopted several approaches, including the GMOR relation, the low-energy theorem plus the LQCD measurement for the topological susceptibility, and the fits to the equations of state from the LQCD$_{\rm T}$ and LQCD$_{\rm I}$ calculations.
The QCD vacuum energy is constrained within a small range of $(163~{\rm MeV})^4< \mbox{QCD VE}<(190~{\rm MeV})^4$. 
Based on our findings and assuming the validity of the $\mathcal{O}(g^4)$ pQCD calculations, we have excluded the stable existence of completely deconfined, zero gluon and quark condensation 2+1- and 2-flavor quark nugget, while for the more sophisticated quark nugget systems, one will need to turn to other approaches beyond this paper to know its stability.

\vspace{1cm}
\subsubsection*{Acknowledgments}
We thank Bob Holdom for useful discussion. This work is supported by the U.S. Department of Energy under the contract DE-SC0017647. YB is also supported by the U.S. Department of Energy under contracts No. DE-AC02-06CH11357 at Argonne National Laboratory. TKC is also supported by the Ministry of Education, Taiwan, under the Government Scholarship to Study Abroad. 

\begin{appendix}

\section{HpQCD formulas for the quark-gluon plasma pressure}\label{sec:HpQCD}

We recap in this section the formulas of HpQCD for the pressure of the quark-gluon plasma at a finite temperature $T$ from Refs.~\cite{Kajantie:2002wa,Laine:2006cp}. For a system of $N_f$ massless quark flavors with $N_c=3$ and the renormalization scale $\overline{\mu}$, the pressure $p_{\rm QCD,0}(\overline{\mu},T,\Delta)$ is given up to $\mathcal{O}(g^6\ln g)$ by
\begin{equation}
    p_{\rm QCD,0}(\overline{\mu},T,\Delta) = \frac{8\pi^2}{45}T^4\left[\sum_{i=0}^6\left(\frac{\alpha_s(\overline{\mu})}{\pi}\right)^{i/2}p_i\right] ~,
\end{equation}
where
\begin{align}
    p_0 &= 1+\frac{21}{32}N_f ~, \\
    p_1 &= 0 ~, \\
    p_2 &= -\frac{15}{4}\left(1+\frac{5}{12}N_f\right) ~, \\
    p_3 &= 30\left(1+\frac{1}{6}N_f\right)^{3/2} ~, \\
    p_4 &= 237.2 + 15.96N_f - 0.4150N_f^2 + \frac{135}{2}\left(1+\frac{1}{6}N_f\right)\ln\left[\frac{\alpha_s}{\pi}\left(1+\frac{1}{6}N_f\right)\right] \nonumber \\
    & \quad- \frac{165}{8}\left(1+\frac{5}{12}N_f\right)\left(1-\frac{2}{33}N_f\right)\ln\frac{\overline{\mu}}{2\pi T}~, \\
    p_5 &= \left(1+\frac{1}{6}N_f\right)^{1/2}\bigg[-799.1-21.96N_f-1.926N_f^2 \nonumber \\
    & \quad+ \frac{495}{2}\left(1+\frac{1}{6}N_f\right)\left(1-\frac{2}{33}N_f\right)\ln\frac{\overline{\mu}}{2\pi T}\bigg] ~, \\
    p_6 &= \bigg[ -659.2-65.89N_f-7.653N_f^2 \nonumber \\
    & \quad +\frac{1485}{2}\left(1+\frac{1}{6}N_f\right)\left(1-\frac{2}{33}N_f\right)\ln\frac{\overline{\mu}}{2\pi T}\bigg]\ln\left[\frac{\alpha_s}{\pi}\left(1+\frac{1}{6}N_f\right)\right] \nonumber \\
    & \quad -475.6\ln\frac{\alpha_s}{\pi} - \frac{1815}{16}\left(1+\frac{5}{12}N_f\right)\left(1-\frac{2}{33}N_f\right)^2\ln^2\frac{\overline{\mu}}{2\pi T} \nonumber \\
    & \quad + (2932.9+42.83N_f-16.48N_f^2+0.2767N_f^3)\frac{\overline{\mu}}{2\pi T} + \Delta(N_f) ~,
\end{align}
with $\Delta$ being an unknown constant to be fit. In our study, we define the renormalization scale parameter $X_T\equiv\overline{\mu}/(2\pi T)$. We consider the running of $\alpha_s$ and $m_s$ up to $\mathcal{O}(\alpha_s^2)$,
\beqa
    \alpha_s(\overline{\mu}) &=& \frac{4\pi}{\beta_0L}\left(1-\frac{2\beta_1}{\beta_0^2}\frac{\ln L}{L}\right) ,~\qquad \mbox{with}\quad L = \ln\left(\overline{\mu}^2/\Lambda_{\overline{\rm MS}}^2\right) ~, \\
    m_s(\overline{\mu}) &=& m_s(2\,{\rm GeV})\left(\frac{\alpha_s(\overline{\mu})}{\alpha_s(2\,{\rm GeV})}\right)^{\gamma_0/\beta_0}\times\frac{1+A_1\frac{\alpha_s(\overline{\mu})}{\pi}+\frac{A_1^2+A_2}{2}\left(\frac{\alpha_s(\overline{\mu})}{\pi}\right)^2}{1+A_1\frac{\alpha_s(2\,{\rm GeV})}{\pi}+\frac{A_1^2+A_2}{2}\left(\frac{\alpha_s(2\,{\rm GeV})}{\pi}\right)^2} ~,
\eeqa
where we choose $\Lambda_{\overline{\rm MS}}=378$~MeV, $m_s(2\,{\rm GeV}) = 93.5\,\mbox{MeV}$, and
\begin{equation}
    A_1 = -\frac{\beta_1\gamma_0}{2\beta_0^2} + \frac{\gamma_1}{4\beta_0} ,\qquad A_2 = \frac{\gamma_0}{4\beta_0^2}\left(\frac{\beta_1^2}{\beta_0}-\beta_2\right) - \frac{\beta_1\gamma_1}{8\beta_0^2} + \frac{\gamma_2}{16\beta_0} ~.
\end{equation}
The $SU(3)_C$ group theory factors as well as the $\beta_i$ and $\gamma_j$ factors are summarized as follows:
\beqa
&&    d_A = N_c^2 - 1 ~, \qquad C_A = N_c ~, \qquad C_F = \frac{N_c^2-1}{2N_c} ~, \nonumber \\
&&    \beta_0 = \frac{11C_A-2N_f}{3} ~,\qquad \qquad \beta_1 = \frac{17}{3}C_A^2 - C_FN_f - \frac{5}{3}C_AN_f  ~,  \nonumber \\
&&    \beta_2 = \frac{2857}{216}C_A^3 + \frac{1}{4}C_F^2N_f - \frac{205}{72}C_AC_FN_f - \frac{1415}{216}C_A^2N_f + \frac{11}{36}C_FN_f^2 + \frac{79}{216}C_AN_f^2 ~, \nonumber \\
&&    \gamma_0 = 3C_F ~, \qquad \qquad \qquad \gamma_1 = C_F\left(\frac{97}{6}C_A+\frac{3}{2}C_F-\frac{5}{3}N_f\right) ~,  \nonumber \\
&&    \gamma_2 = C_F\Bigg\{\frac{129}{2}C_F^2-\frac{129}{4}C_FC_A+\frac{11413}{108}C_A^2+C_FN_f\left[-23+24\zeta(3)\right] \nonumber \\
&&   \hspace{1.5cm} + \, C_AN_f\left[-\frac{278}{27}-24\zeta(3)\right]-\frac{35}{27}N_f^2\Bigg\} ~. \label{eq:QCD:group}
\eeqa

The massive quark effects are only included up to $\mathcal{O}(g^2)$ in Ref.~\cite{Laine:2006cp} by modifying $p_{\rm QCD,0}$ by
\begin{equation}
    p_{\rm QCD,m}(N_f,\overline{\mu},T,\Delta)  \equiv \frac{\left[\alpha_{E1}^{\overline{\rm MS}}+g^2\alpha_{E2}^{\overline{\rm MS}}\right](N_f)}{\left[\alpha_{E1}^{\overline{\rm MS}}+g^2\alpha_{E2}^{\overline{\rm MS}}\right](0)}\times p_{\rm QCD,0}(\overline{\mu},T,\Delta) ~,
\end{equation}
where
\begin{align}
    \alpha_{E1}^{\overline{\rm MS}} &= d_A\frac{\pi^2}{45} + 4C_A\sum_{i=1}^{N_f} F_1\left(\frac{m_i^2}{T^2}\right) ~, \\
    \alpha_{E2}^{\overline{\rm MS}} &= -\frac{d_AC_A}{144} - d_A\sum_{i=1}^{N_f}\bigg\{ \frac{1}{6}F_2\left(\frac{m_i^2}{T^2}\right)\left[1+6F_2\left(\frac{m_i^2}{T^2}\right)\right] \nonumber \\
    & \quad +\frac{m_i^2}{4\pi^2 T^2}\left(3\ln\frac{\overline{\mu}}{m_i}+2\right)F_2\left(\frac{m_i^2}{T^2}\right) - \frac{2m_i^2}{T^2}F_4\left(\frac{m_i^2}{T^2}\right)\bigg\} ~,
\end{align}
with
\begin{align}
    F_1(y) &= \frac{1}{12\pi^2}\int_0^\infty dx\left(\frac{x}{x+y}\right)^{1/2}n_F(\sqrt{x+y})x ~, \\
    F_2(y) &= \frac{1}{4\pi^2}\int_0^\infty dx \left(\frac{x}{x+y}\right)^{1/2}n_F(\sqrt{x+y}) ~, \\
    F_4(y) &= \frac{2}{(4\pi)^4}\int_0^\infty dx_1\int_0^\infty dx_2\frac{1}{\sqrt{x_1+y}\sqrt{x_2+y}} n_F(\sqrt{x_1+y})n_F(\sqrt{x_2+y}) \nonumber \\
    &\times \ln\left[ \frac{\sqrt{x_1+y}\sqrt{x_2+y}+y-\sqrt{x_1x_2}}{\sqrt{x_1+y}\sqrt{x_2+y}+y+\sqrt{x_1x_2}}\times \frac{\sqrt{x_1+y}\sqrt{x_2+y}-y+\sqrt{x_1x_2}}{\sqrt{x_1+y}\sqrt{x_2+y}-y-\sqrt{x_1x_2}} \right] ~,
\end{align}
and $n_F(x)=1/(e^x+1)$. Finally, we include the bag parameter $B$ and obtain the total pressure
\begin{equation}
    p_{\rm QCD}(N_f,\overline{\mu},T,\Delta,B) = p_{\rm QCD,m}(N_f,\overline{\mu},T,\Delta) - B ~.
\end{equation}

\section{The validity of \texorpdfstring{$\mathcal{O}(g^4)$~}~pQCD calculations}\label{sec:NLO}

\begin{figure}[ht!]
    \centering
    \includegraphics[width=0.49\linewidth]{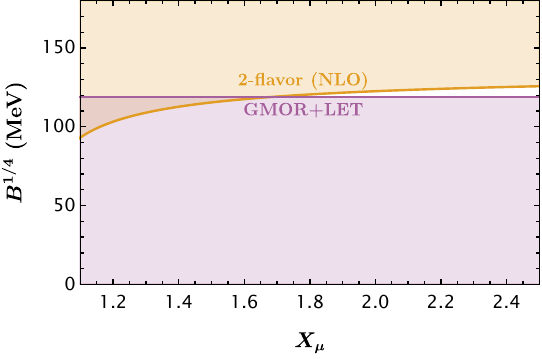}
    \includegraphics[width=0.49\linewidth]{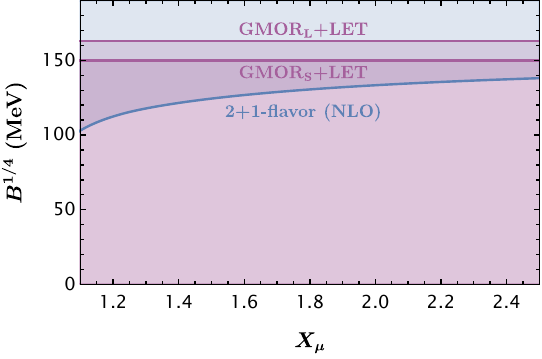}
    \caption{Similar plots to Figures~\ref{fig:X_bag_2-main} and \ref{fig:X_bag_3-main} but with pQCD only up to NLO ($\mathcal{O}(g^2)$). 
    }
    \label{fig:stability_NLO}
\end{figure}

In this section, we discuss the validity of pQCD at $\muB\approx930$~MeV and present the quark nugget stability analysis based on the possible scenario where the $\mathcal{O}(g^6)$ correction completely cancels the $\mathcal{O}(g^4)$ correction, leaving only the $\mathcal{O}(g^2)$ correction intact. As we presented in Ref.~\cite{Bai:2024amm}, both the $\mathcal{O}(g^2)$ and $\mathcal{O}(g^4)$ corrections tend to reduce the quark matter pressure and are justifiably small compared to the $\mathcal{O}(g^0)$ contribution.
From the partial $\mathcal{O}(g^6)$ results presented in Refs.~\cite{Gorda:2021znl,Gorda:2021kme,Gorda:2023mkk}, this reduction in the quark matter pressure seems to be further strengthened, which thus further solidifies the exclusion of the certain kind of stable quark nugget considered in this study.
Furthermore, the perturbative expansion parameter $[\alpha_s(\muB)/\pi]<1$ throughout the $\muB$ range of our interest around the proton mass, and hence our calculations based on pQCD should be valid for the specific question of quark nugget stability. However, the color $SU(3)_C$ group factors showing up at different loop levels could complicate the convergence of the expansion series, which, in the extreme case, might lead to the complete cancellation between the $\mathcal{O}(g^4)$ and $\mathcal{O}(g^6)$ corrections, though it seems highly unlikely according to Refs.~\cite{Gorda:2021znl,Gorda:2021kme,Gorda:2023mkk}. To investigate this situation, we perform a similar analysis as in the main text but based on the $\mathcal{O}(g^2)$ (NLO) framework to study the stability of the 2- and 2+1-flavor quark nuggets, the results of which are shown in Figure~\ref{fig:stability_NLO}. As can be seen from the plots, the exclusion of the stable 2+1-flavor quark nugget remains concrete, while for the 2-flavor case, there is a chance that quark nugget can be stable for $X_\mu\gtrsim1.7$. However, it is highly inconceivable that the convergence of pQCD will be so poor for such large renormalization scales close to the usual maximum value used in the literature ($X_\mu=2.0$) since the cutoff baryon chemical potential $\mu_0$ is way smaller than $930$~MeV at $\mathcal{O}(g^4)$ (see Ref.~\cite{Bai:2024amm} for details). Also, as we mentioned in footnote [32], our estimation of $B_G^{\rm con}$ using the LET is conservative in the sense that we do not consider the $\mathcal{O}(m_q)$ corrections, which, if taken into account, should in principle make the exclusion more robust. Nevertheless, as we brought up earlier, there is no guarantee at this point that the $\mathcal{O}(g^6)$ correction will not reverse the conclusions of this study until the full calculations proposed in Ref.~\cite{Karkkainen:2025nkz} are carried out.
\end{appendix}
\setlength{\bibsep}{3pt}
\providecommand{\href}[2]{#2}\begingroup\raggedright\endgroup

\end{document}